\documentclass[prd, superscriptaddress,nofootinbib,twocolumn]{revtex4}
\usepackage{graphicx,natbib}
\usepackage{url}
\usepackage{color}
\usepackage{rotating}
\usepackage{amssymb,amsmath}

\def\hmpc{\,h^{-1}\,{\rm Mpc}}
\def\Mpc{{\rm Mpc}}
\def\Gpc{{\rm Gpc}}
\newcommand{\fnl}{f_{\rm NL}}
\newcommand{\fnleq}{f_{\rm NL}^{\rm eq}}
\newcommand{\Lpix}{L_{\rm pix}}
\newcommand{\thetapix}{\theta_{\rm pix}}

\newcommand{\degr}{\hbox{$^\circ$}}
\newcommand{\sqdeg}{\,{\rm deg}^2}
\newcommand{\zmax}{z_{\rm max}}

\newcommand{\Mobs}{M_{\rm obs}}
\newcommand{\Mbias}{M^{\rm bias}}
\newcommand{\Mth}{M^{\rm th}}
\newcommand{\siglnM}{\sigma_{\ln M}}
\newcommand{\zphot}{z^{\rm p}}
\newcommand{\zbias}{z^{\rm bias}}
\newcommand{\sigz}{\sigma_{z}}
\newcommand{\Msun}{M_{\odot}}
\newcommand{\DE}{\Omega_{\rm DE}}

\begin{document}

\title{Primordial non-Gaussianity from the covariance of galaxy cluster counts}

\author{Carlos Cunha}
\affiliation{Department of Physics, University of Michigan, 
450 Church St, Ann Arbor, MI 48109-1040}

\author{Dragan Huterer}
\affiliation{Department of Physics, University of Michigan, 
450 Church St, Ann Arbor, MI 48109-1040}

\author{Olivier Dor\'{e}}
\affiliation{
Jet Propulsion Laboratory, California Institute of Technology,
Pasadena, CA 91109}
\affiliation{
California Institute of Technology, Pasadena, CA 91125}

\date{\today}

\begin{abstract}
  It has recently been proposed that the large-scale bias of dark matter halos
  depends sensitively on primordial non-Gaussianity of the local form.  In
  this paper we point out that the strong scale dependence of the non-Gaussian
  halo bias imprints a distinct signature on the covariance of cluster counts.
  We find that using the full covariance of cluster counts results in
  improvements on constraints on the non-Gaussian parameter $\fnl$ of three
  (one) orders of magnitude relative to cluster counts (counts + clustering
  variance) constraints alone.  We forecast $\fnl$ constraints for the
  upcoming Dark Energy Survey in the presence of uncertainties in the
  mass-observable relation, halo bias, and photometric redshifts. We find that
  the DES can yield constraints on non-Gaussianity of
  $\sigma(\fnl)\sim 1$-$5$ even for relatively conservative
  assumptions regarding systematics. Excess of correlations of cluster counts
  on scales of hundreds of megaparsecs would represent a smoking gun
  signature of  primordial non-Gaussianity of the local type.
\end{abstract}

\maketitle
\section{Introduction}\label{sec:intro}

Primordial non-Gaussianity provides cosmology one of the precious few
connections between primordial physics and the present-day universe. Standard
inflationary theory with a single-field, slowly rolling scalar field, predicts
that the spatial distribution of structures in the universe today is very
nearly Gaussian random (e.g.\
\cite{maldacena,Acquaviva:2002ud,Creminelli:2003iq,Lyth_Rodriguez,Seery_Lidsey};
for an excellent recent review, see \cite{Chen_AA}).  Departures from
Gaussianity, barring contamination from systematic errors or late-time
non-Gaussianity due to secondary processes, can therefore be interpreted as
violation of this ``vanilla'' inflationary assumption.  Constraining or
detecting primordial non-Gaussianity is therefore an important and basic test
of the cosmological model.

Constraints on primordial non-Gaussianity have been traditionally obtained from observations of the
cosmic microwave background, as nonzero non-Gaussianity generates a non-zero
three-point correlation function (or its Fourier transform, the bispectrum) of
density fluctuations
\cite{Falk_Ran_Sre,Luo_Schramm,Gangui_etal,Wang_Kam,Komatsu_Spergel,Bartolo_AA,Liguori_AA}.
Increasingly sophisticated algorithms have been developed to constrain
non-Gaussianity,
\cite{Babich:2005en,Babich_shape,Creminelli_estimators,Smith_Zaldarriaga,Fergusson_Shellard}
and, to the extent that it can be measured, Gaussianity has so far been
confirmed \cite{wmap3,Creminelli_wmap,SenSmiZal09}. For example, the most
recent constraints from the Wilkinson Microwave Anisotropy Probe (WMAP)
indicate $\fnl\approx 32\pm 21$ ($1\sigma$; \cite{wmap7}), where the exact
constraints depend somewhat on the choice of the statistical estimator applied
to the data, the CMB map used, and details of the foreground subtraction. Here
$\fnl$ is the parameter describing non-Gaussianity in the widely studied
``local'' model, where the non-Gaussian potential $\Phi_{\rm NG}$ is defined
by
\begin{equation}
\Phi_{\rm NG}(x)=\Phi_{\rm G}(x) + \fnl (\Phi_{\rm G}^2(x)-\langle\Phi_{\rm
  G}^2\rangle),
\label{eq:fnl_local}
\end{equation}
and where $\Phi_{\rm G}$ is the Gaussian potential.  Corresponding constraints
can be obtained on other classes of non-Gaussian models. For example,
for ``equilateral'' models where most power comes from equilateral triangle
configurations, 
$\fnleq=26\pm 140$
($1\sigma$; \cite{wmap7}). 

The CMB is not the only cosmological probe to be sensitive to the presence of
primordial non-Gaussianity. It has been known for a relatively long time that
the abundance of dark matter halos
\cite{Lucchin:1987yv,Robinson:1999se,Benson:2001hc,MVJ,verde01,sco04,Komatsu:2003fd}
(or voids \cite{Kam_voids,songvoids}) is sensitive to the presence of primordial
non-Gaussianity. This dependence is easy to understand: halos populate the
high tail of the probability density distribution of structures in the
universe, and the shape of this distribution is sensitive to departures from
Gaussianity. However, while the halo abundance is rather powerful in
constraining models that are non-Gaussian in the density (rather than the
potential) \cite{Verde_tests}, for the popular models of the local type (cf.
Eq.~(\ref{eq:fnl_local})) the abundance is much less constraining than the CMB
anisotropy and not competitive with the CMB constraints (e.g.\
\cite{Verde_CMBLSS,Sefusatti}).

Some of us \cite{Dalal} have recently shown that the clustering of dark matter
halos is very sensitive to primordial non-Gaussianity of the local type. This
exciting development paves way to using the large-scale structure to probe
primordial non-Gaussianity nearly three orders of magnitude more accurately
than using the abundance of halos.  \citet{Dalal} found, analytically and
numerically, that the bias of dark matter halos acquires strong scale
dependence
\begin{equation}
b(k)=b_0 + \fnl(b_0-1)\delta_c\, \frac{3\Omega_mH_0^2}{a\,g(a) T(k)c^2 k^2},
\label{eq:bias}
\end{equation}
where $b_0$ is the usual Gaussian bias (on large scales, where it is
constant), $\delta_c\approx 1.686$ is the collapse threshold, $a$ is the scale
factor, $\Omega_M$ is the matter fraction relative to critical, $H_0$ is the
Hubble constant, $k$ is the wavenumber, $T(k)$ is the transfer function, and
$g(a)$ is the growth suppression factor\footnote{The usual linear growth
    $D(a)$, normalized to be equal to $a$ in the matter-dominated epoch, is
    related to the suppression factor $g(a)$ via $D(a)=ag(a)$.}. This result
has been confirmed by other researchers using a variety of methods, including
the peak-background split \cite{Afshordi_Tolley,MV,Slosar_etal}, perturbation
theory \cite{McDonald,Taruya08,GP}, and numerical (N-body) simulations
\cite{Grossi,Desjacques_Seljak_Iliev,PPH}.  Astrophysical measurements of the
scale dependence of the large-scale bias, using galaxy and quasar clustering
as well as the cross-correlation between the galaxy density and CMB
anisotropy, have recently been used to impose constraints on $\fnl$ 
already comparable to those from the cosmic microwave background (CMB)
anisotropy \cite{Slosar_etal, Afshordi_Tolley}, giving $\fnl=28\pm 23$
($1\sigma$), with some dependence on the assumptions made in the analysis
\cite{Slosar_etal}. In the future, constraints on $\fnl$ are expected to be of
order a few \cite{Dalal,Carbone,Carbone2}. The sensitivity of the large-scale
bias to other models of primordial non-Gaussianity has not been investigated
yet (though see preliminary analyses in \cite{Desjacques_gnl,MV09}).

{\it Clustering} of galaxy clusters, in particular, can very strongly
constrain primordial non-Gaussianity.  Clusters have an advantage of being
large, relatively simple objects that are easy to find using either optical or
X-ray light, or else from their Sunyaev-Zeldovich signature. Clusters already
provide interesting constraints on dark energy \cite{Mantz07,Vikhlinin} and
they hold promise for precision measurements of cosmological and dark energy
parameters (e.g.\ \cite{clusters_PC}). Since clusters are massive and hence
significantly biased objects, their counts (via the mass function) and
clustering (via the mass function and bias) are both sensitive to primordial
non-Gaussianity.  Recently, \citet{Oguri09} has argued that the variance of
cluster counts (i.e.\ scatter measured in each cell individually), 
in combination with the cluster counts, leads to interesting improvements
on $\fnl$ constraints relative to the counts-only case.

In this paper we point out that including the {\it covariance} of cluster
counts in angle and redshift leads to very significant further improvements in
the cluster constraints on local primordial non-Gaussianity.  The principal
reason for the improvement is simple: covariance is determined by the cluster
power spectrum, which is proportional to the halo bias squared.  At large
scales, the non-Gaussian contribution to the halo bias dominates
(cf.\ Eq.~(\ref{eq:bias})), and this results in a strong $\fnl$ signal in the
covariance.  Furthermore, we explore the sensitivity of the constraints to
various assumptions about statistical and systematic errors in modeling the
cluster mass-observable relation, as well as the presence of other
cosmological parameters.  We find that the bulk of the information about local
non-Gaussianity comes from the far-separation covariances of cluster
counts-in-cells.

This paper is organized as follows. In Sec.~\ref{sec:method}, we describe the
methodology that we use to obtain constraints from both counts and clustering
of galaxy clusters. In Sec.~\ref{sec:calc} we describe our fiducial
assumptions about the cosmological model and data as well as solutions to
various challenges in calculating the constraints. In Sec.~\ref{sec:results}
we describe the forecasted constraints on $\fnl$ from the Dark Energy
Survey. We discuss our results in Sec.~\ref{sec:disc}, and conclude in Sec.~\ref{sec:concl}.

\section{Methodology}\label{sec:method}

We address the following problem: how well can the cosmological parameters be
recovered using counts of galaxy clusters in pixels distributed  in angle and
radius on the sky? We largely follow the formalism of \citet{Hu_Cohn} and
\citet{lim05}.

Assume that clusters are counted in square pixels of fixed angular size
$\thetapix$, corresponding to comoving size $\Lpix(z) = \thetapix r(z)$, where
$r$ is the comoving distance. The clusters are also binned in the mass-observable 
(i.e the observable proxy for cluster mass), with
intervals $[\Mobs^{\alpha}, \Mobs^{\alpha+1}]$ where $\alpha$  refers to a
specific mass-observable bin. 
The number density of clusters at a given redshift $z$
with observable in the range $\Mobs^{\alpha} \leq \Mobs \leq \Mobs^{\alpha+1}$
is given by
\begin{eqnarray}
\bar n_{\alpha}(z) &\equiv& \int_{\Mobs^{\alpha}}^{\Mobs^{\alpha+1}} \frac{d \Mobs}{\Mobs}
\int {\frac{dM}{M}} { \frac{d \bar n}{d\ln M}}
p(\Mobs | M)
\label{eqn:nofz}
\end{eqnarray}
\noindent where $p(\Mobs | M)$ is the observable-mass relation (explained in
Appendix \ref{app:mobs}) and $d \bar n/d\ln M$ is the mass function.
Uncertainties in the redshifts distort the volume element; we fully take into
account the photometric redshift uncertainties following \cite{lim07}; details
are shown in Appendix \ref{app:photoz}. 

We adopt the mass function from \citet{Dalal} who used N-body simulations to
parametrize the shift in mass of a typical halo in the presence of non-Gaussianity.
The mass shift, $M_{\rm G}\rightarrow M$, is adequately described by a
Gaussian with mean and variance respectively given by
\begin{eqnarray}
\left\langle\frac{M}{M_{\rm G}}\right\rangle -1
& = & 1.3\times 10^{-4}\,\fnl \sigma_8\,\sigma(M_{\rm G},z)^{-2} \label{eq:mean_Mf_def}
\\[0.1cm]
{\rm var}\left(\frac{M}{M_{\rm G}}\right)
& = & 1.4\times 10^{-4}\, (\fnl\sigma_8)^{0.8} \sigma(M_{\rm G},z)^{-1},
\label{eq:rms_Mf_def}
\end{eqnarray}
where $\sigma(M,z)$ is the amplitude of mass fluctuations on mass scale $M$
and at redshift $z$. The final non-Gaussian mass function is given by \cite{Dalal}
\begin{equation}
\frac{dn}{dM} = \int dM_{\rm G} \frac{dn}{dM_{\rm G}} \frac{dP}{dM}(M_{\rm G}),
\label{eq:mf_conv}
\end{equation}
where $dP/dM(M_{\rm G})$ is the probability distribution that a Gaussian halo of
mass $M_{\rm G}$ maps to a non-Gaussian halo of mass $M$, and is given by the
Gaussian with the mean and variance given in Eqs.~(\ref{eq:mean_Mf_def}) and
(\ref{eq:rms_Mf_def}). For $dn/dM_{\rm G}$, we adopt the Jenkins mass function
\cite{jenkins}.

On large scales, the number counts of clusters $m({\bf{x}})$ trace
the linear density perturbation $\delta({\bf{x}})$
\begin{equation}
m_i(M_\alpha, {\bf x}) \equiv m_{i\alpha}= \bar{m}_i ( 1 + b(M_\alpha, z)\delta({\bf x}))
\label{eqn:count_fluc}
\end{equation}
where $i$ refers to the pixel (i.e.\ its angular and radial coordinates), and
$\alpha$ indicates the mass bin.  The spatial covariance of cluster counts is
\cite{HuKra03}
\begin{equation}
S^{\alpha}_{ij} = \langle (m_{i\alpha} -\bar m_{i\alpha})(m_{j\alpha}
- \bar m_{j\alpha}) \rangle 
\equiv  \bar m_{i\alpha} \bar m_{j\alpha} 
\xi^{\alpha}_{ij}, 
\label{eq:Sij}
\end{equation}
where $\xi^{\alpha}_{ij} $ is the pixel real-space correlation function
\begin{eqnarray}
\xi^{\alpha}_{ij} &\equiv & \int \frac{d^3 k}{(2\pi)^3} |W_i({\bf k})W_j({\bf k})|
\cos(k_x \Delta x_{ij})\times \nonumber \\[0.1cm]
& & \cos(k_y \Delta y_{ij})\cos(k_z \Delta z_{ij})
b_{i\alpha}b_{j\alpha} P(k;z).
\label{eq:Sij_bare}
\end{eqnarray} 
\noindent  
If $i$ and $j$ come from different redshift bins, the geometric mean
of the two redshifts\footnote{
  In the linear regime, the correlation between pixels
  $i$ and $j$ contains the product of the growth factors corresponding to
  $z_i$ and $z_j$.  Therefore, the corresponding power spectrum, $P(k, z)$ in
  Eq.~(\ref{eq:Sij_bare}), should use the growth function equal to the
  geometric mean of the two growth functions.  Instead, we effectively use the
  growth function which is evaluated at the redshift equal to the geometric
  mean of the two {\it redshifts} $z_i$ and $z_j$.  Results are
  insensitive to this approximation, specially because most of 
  the information comes from relatively close redshift pairs.
}  is adopted for $z$.  In the limit $r_{ij} \gg L_{\rm pix}$,
$\xi_{ij}^\alpha \to \xi(r_{ij})$, where the latter quantity is the standard
two-point correlation function in real space.  $\Delta x_{ij} = L_{\rm pix}
n_{xij}$ is the physical separation between $i$ and $j$ in the $x$ direction
(transverse to the line of sight), and $n_{xij}$ is the number of pixels
separating them; $\Delta y_{ij}$ is defined equivalently.  Finally, the window
function $W$ is the Fourier transform of the square pixel in the presence of
redshift errors
\begin{eqnarray}
W({\bf k})_i &=& 
\exp\left (\frac{-\sigma^{2}_{z,i}k^2_z}{2 H^2_i}\right  ) \times  \\[0.1cm] 
&&j_0(k_x L_{\rm pix} /2) j_0(k_y L_{\rm pix} /2) j_0(k_z \Delta z/2H_i) \,, \nonumber
\end{eqnarray}
where the index $i$ refers to the redshift bin, $\sigma_{z,i}$ is the redshift
scatter at the radial distance corresponding to the $i$th pixel, and $H_i$ is
the Hubble parameter.  The photo-z bias is implicit in the $\Delta z_{ij}$
term in Eq.~(\ref{eq:Sij_bare}).

The expression for the full Fisher matrix for galaxy cluster counts
and their covariance is quite complicated (see \cite{Hu_Cohn}), but a reasonable
approximation is given by \citep{lim04}
\begin{eqnarray}
F_{\mu\nu}&=&  \bar{\bf m}^t_{,\mu} {\bf C}^{-1}
 \bar{\bf m}_{,\nu} 
+ \frac{1}{2} {\rm Tr} [{\bf C}^{-1} {\bf S}_{,\mu}
 {\bf C}^{-1} {\bf S}_{,\nu} ]\,,
 \label{eq:fisher}
\end{eqnarray}
where the first term encodes information from cluster counts, and the second
from the covariance. Here $\mu$ and $\nu$ are indices that refer to both
cosmological and nuisance parameters (including $\fnl$). The cluster counts
have been arranged as the vector $\bar{\bf m}$. ${\bf S}=\{S^{\alpha}_{ij}\}$ is
the sample covariance matrix from Eq.~(\ref{eq:Sij}), and ${\bf C} \equiv {\bf
  N + S}$ is the total covariance. $N_{ij}=\bar m_i\delta_{ij}$ is the (shot) noise matrix.  The
derivative with respect to $\fnl$ can be computed analytically, using the fact
that $P(k, z)\propto b^2(k, z)$ and Eq.~(\ref{eq:bias}).

\begin{figure*}[!t]
\includegraphics[scale=0.4]{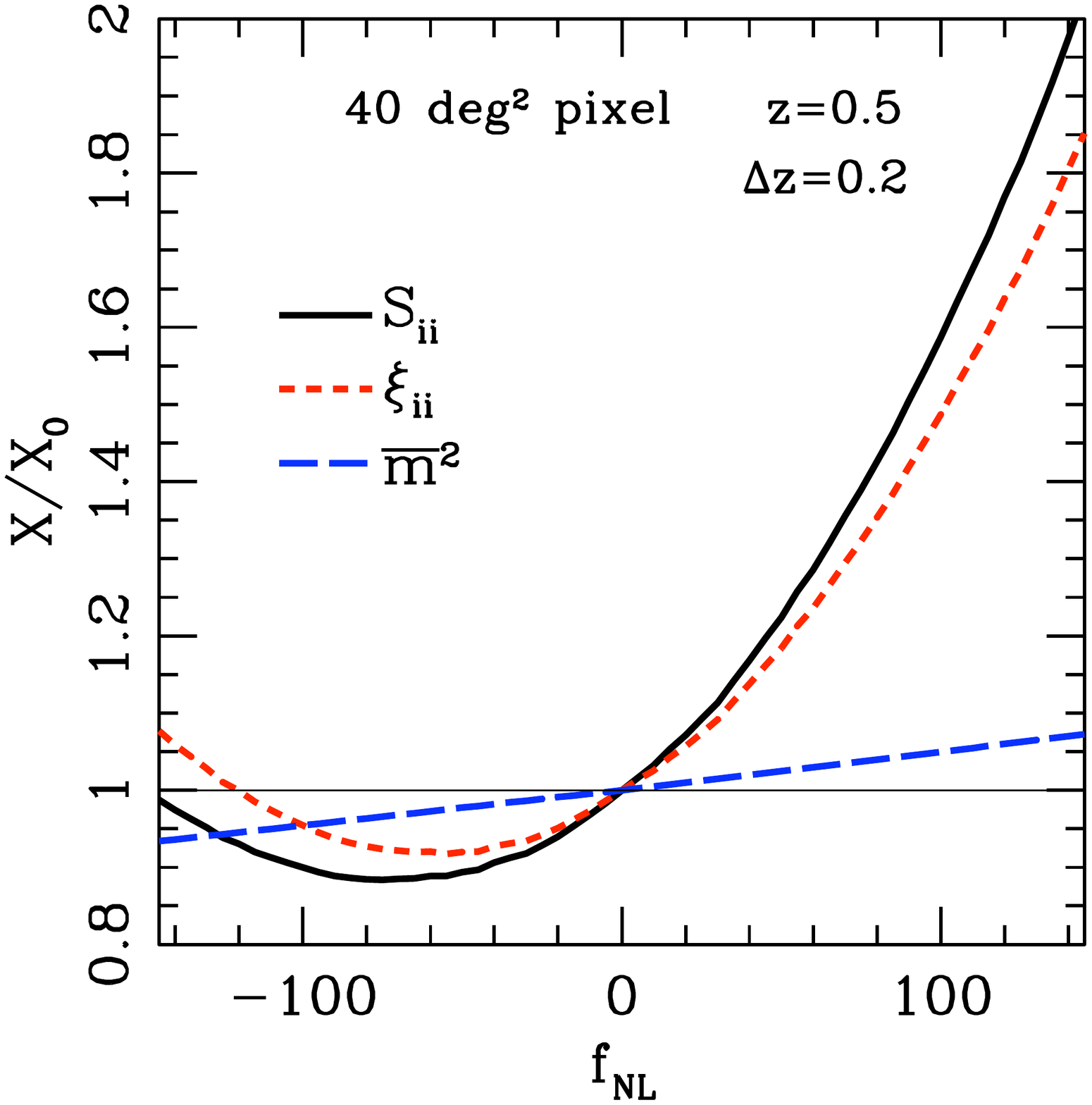}\hspace{0.5cm}
\includegraphics[scale=0.4]{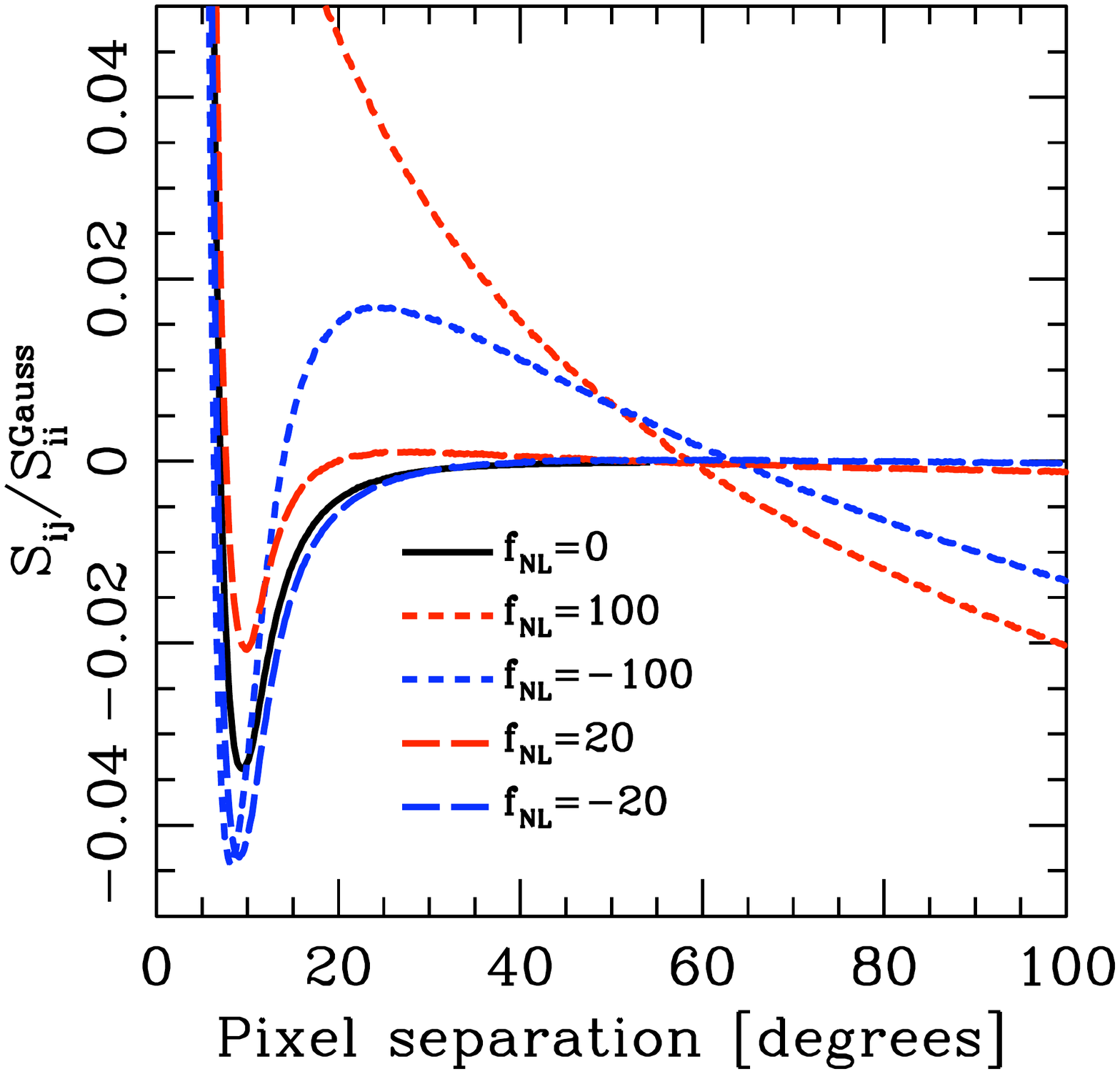}
\caption{{\it Left panel}: Sensitivity of the variance of cluster counts to
  non-Gaussianity.  The black lines shows the variance $S_{ii}$, the
  short-dashed red line shows the (auto)correlation function $\xi_{ii}^{\alpha}$,
  and the long-dashed blue line shows the squared mean counts.  Note that
  $S_{ii}=({\bar m})^2\xi_{ii}^{\alpha}$ We assumed a pixel with area
  40 sq.\ deg.\ and radial redshift extent $\Delta z=0.2$, centered at
  $z=0.5$. {\it Right panel}: Sensitivity of the covariance of cluster counts to
  non-Gaussianity. We show the off-diagonal elements of the clustering matrix,
  normalized by the variance of $\fnl=0$ case ($S_{ii}^{\rm Gauss}$) 
  as a function of angle between the $i$th and $j$th pixel.
  We use the same pixelization as for the left panel.
  We show the Gaussian case ($\fnl=0$), and four non-Gaussian models
  ($\fnl=\pm 20$ and $\fnl=\pm 100$). 
  Note that, because of the regularization, the results depend on the size of the survey.
  The larger the survey, the larger the effect of non-Gaussianity.
}
\label{fig:Sij}
\end{figure*}

\section{Fiducial assumptions and Calculational challenges}\label{sec:calc}

We implement the procedure outlined above for optically selected clusters.  In
our fiducial setup we divide the sky into the $11\times11$ field of pixels of
$41.32$ sq.\ deg.\ each, for a total of 5,000 sq.\ deg.\ which matches
expectations for the Dark Energy Survey (DES). The facing surface of each
pixel is a square with a side $\Lpix(z) = \thetapix r(z)$ (see
Sec.~\ref{sec:method}).  Each pixel has redshift depth $\Delta z=0.2$, and we
assume a maximum redshift of 1.0 so that there are five redshift bins. We
adopt the mass threshold of $\Mth=10^{13.7} h^{-1}\Msun$ and also bin in mass,
using 5 mass bins of width $\Delta \ln \Mth = 0.2$, with the exception of the
highest-mass bin, which we extend to infinity.  Using smaller bins in angle or redshift
yields better results, up to the point where the covariance matrix becomes
dominated by shot-noise (which occurs for bins with area around 0.1-1
sq. deg.).  For very large number of pixels, the Gaussian approximation used
to define the covariance used in our Fisher matrix would break down.  In our
fiducial case we have about $1.7\times10^5$ clusters subdivided into $3,025$
pixels, so that we are well within the Gaussian regime.  In addition, results
for large angular pixels are less sensitive to systematics due to non-linear
physics or angular mask uncertainties.  In Sec. \ref{sec:results} we consider
departures from the fiducial assumption, namely variations in the
mass-threshold, maximum redshift range and pixel area.

We assume fiducial cosmological parameters based on the fifth year data
release of the Wilkinson Microwave Anisotropy Probe \citep{wmap5}.  Thus, we
set the baryon density, $\Omega_b h^2=0.0227$, the dark matter density,
$\Omega_m h^2 =0.1326$, the normalization of the power spectrum at $k=0.05
{\rm Mpc}^{-1}$, $\delta_{\zeta}=4.625 \times 10^{-5}$, the tilt, $n=0.963$,
the optical depth to reionization, $\tau=0.087$, the dark energy density,
$\Omega_{\rm DE}=0.742$, and the dark energy equation of state, $w=-1$.  In
this cosmology, $\sigma_8=0.796$.  We use CMBfast \citep{sel96}, version
4.5.1, to calculate the transfer functions, and add Planck priors\footnote{W.\ Hu,
private communication} when calculating the marginalized constraints on
parameters.

To study systematic errors in cluster cosmology, we add a generous set of
nuisance parameters described in Appendix \ref{app:mobs} (see also
\citet{cun08} and \citet{clusters_PC}), with 10 nuisance parameters describing
the bias and variance in the mass-observable relation and 3 parameters
describing uncertainty in the halo bias ($a_c$, $p_c$, and $\delta_c$,
cf. Eq.~(\ref{eqn:bias})).  The assumption of 3 nuisance parameters describing
the Gaussian halo bias is somewhat {\it ad hoc} 
but conservative since for a given mass function the halo bias can be predicted
to roughly $10\%$ accuracy in the range of scales we are interested
\cite{manera09}.  We fix the photo-z scatter to $0.02$ everywhere except in
Sec.~\ref{sec:photoz} where we consider the effects of including 10 additional
nuisance parameters describing photometric redshift errors.  In this
exploratory paper we do not consider models for non-Gaussianity other than the
one from Eq.~(\ref{eq:fnl_local}), or observational systematic errors
(e.g.\ atmospheric blurring or completeness variations across the sky). The
study of these effects is left for future work.

Evaluating the expression for the Fisher matrix with the signal matrix of this
size is clearly challenging: the total size of the matrix $S$ (see
Eq.~(\ref{eq:Sij_bare})) is $N\times N$, where $N=N_{\rm pixels}\times N_{\rm
  mass}\times N_{\rm redshift}=121\times 5\times 5 = 3,025$ in our fiducial
case.  The bottleneck is in calculating the $\sim 10^7$ elements of the
matrix, each of which involves the numerical computation of a rapidly
oscillating triple integral; see Eqs.~(\ref{eq:Sij}) and (\ref{eq:Sij_bare}).
Unlike previous works which studied constraints on dark energy
\citep{Hu_Cohn,lim04,lim07}, we cannot ignore the off-diagonal elements (i.e.\
the pixel {\it co}variance) of the matrix $S$ since those elements, while
being very small for the Gaussian case, become significant for $\fnl\neq 0$
(see the right panel of Fig.~\ref{fig:Sij}) due to the $\fnl^2 k^{-4}$
dependence scaling of the power spectrum as $k\rightarrow 0$.  To reduce the
size of the covariance matrix we assume that the information from the
different mass bins is independent, so that we can estimate the Fisher matrix
for each mass bin separately and then add them in the end.
The scatter in the mass-observable relation can generate correlations between mass bins
at a given pixel.
In addition, as \citet{Seljak09} and  \citet{McDonald_Seljak09} noticed,
correlating the halos of different masses at large separations 
would lead to improved constraints in our analysis, making our assumption
conservative.  

\subsection{Regularization of the covariance}
As \citet{wands_slosar} pointed out, the two-point correlation function for
biased tracers of structure has an infrared divergence if $\fnl$ is not zero.
However, the {\it measured} correlation function from any survey is of course
finite because one cannot measure variance of the density field on scales
larger than the survey.  To that effect, \citet{wands_slosar} suggest
regularizing the correlation function $\xi(r)$ by subtracting from it the
variance of the density field evaluated at the scale of the survey.  However,
Cunha and Slosar (private communication) found out that the regularization of
 \citet{wands_slosar} contains a typo; the correct regularization term is given by
\begin{equation}
\Sigma^2(R) \equiv
\int \frac{d^3 k}{ (2\pi)^3} |W_R({\bf k})|b_{i\alpha}b_{j\alpha} P(k;z),
\label{eq:var}
\end{equation}
where we use the mass bin $\alpha$ and redshift bin $i$ corresponding to
those of the correlation function $\xi^\alpha_{ij}$ from which this is being
subtracted. If $i$ and $j$ come from different redshift bins,
the geometric mean of the two redshifts is taken.
The difference from what is presented in \citet{wands_slosar} is that our 
expression has $|W_R({\bf k})|$ instead of $|W_R({\bf k})|^2$ (cf. Eqs. 47, 49 and 50 in \cite{wands_slosar}).
Using the above expression, the observed 2-pt correlation at a given survey
volume has the desirable property that it integrates to zero over the 
survey volume.

We approximate the window function $|W_R({\bf k})|$ as the Fourier transform
of a spherical top-hat, and adopt $R=2 h^{-1}\Gpc$ as the linear dimension of
our survey.  For the main analysis in this paper, the effects of the
divergence are not significant, since all of our results (except in Sec.\
\ref{sec:disc}) assume zero fiducial $\fnl$, and the analytic expression for
the derivative $dS_{ij}/d\fnl$ is weakly sensitive to the integration
boundary.  The divergence of the two-point correlation does affect the
covariance for non-zero $\fnl$ and for pixel separation greater than a few
hundred Mpc.  We use the lower boundary of integration $k_{\rm min}=10^{-4}$,
and check that results are stable vis-a-vis variations in this value, or
whether the regularization mentioned above has been applied or not. For
Fig. \ref{fig:Sij} and the results in Sec.~\ref{sec:disc}, we do apply the
corrected Wands-Slosar regularization prescription (cf. Eq. \ref{eq:var}).

Besides its impact on the regularization, the choice of survey geometry is 
important since the {\it distribution} of
pixel-pixel separations depends on the geometry.  We assume that the survey
itself has square shape (and implicitly work in a flat-sky approximation), and
assume a $11\times 11$ field of square-shaped pixels for each redshift bin.
To populate the covariance matrix, we precompute $S_{ij}$ as a function of
pixel separation for integer values of the separation along a row of pixels in
Eq.~(\ref{eq:Sij_bare}) --- that is, we set $\Delta y_{ij}=0$ and vary $\Delta
x_{ij}$ at each redshift. We use linear interpolation to estimate the
covariance for pixels whose physical separation, in units of $\Delta x_{i
  (i+1)}$, is non-integer.  We find that the effects of disregarding the pixel
orientation are negligible (by changing the orientation of bins and finding
little change in the results). Pre-computation of the covariance matrix
elements as a function of pixel separation greatly reduces the number of
covariance terms we need to calculate.

As the right panel of Fig.~\ref{fig:Sij} shows, in the Gaussian case the
off-diagonal terms of $S_{ij}$ fall off very fast.   
We find that covariance terms for pixels in different redshifts to be negligible, 
because we use broad redshift bins.  
Hence, we only calculate covariance between different redshift
bins when estimating the derivative of the covariance with respect to $\fnl$.
To save time, for the results shown in \ref{sec:results} we only calculate
terms in adjacent redshift bins. We checked that including larger redshift
separations improves unmarginalized constraints by about 30\%. But including
the regularization removes most of the improvement (for fiducial $\fnl=0$). To calculate
the derivatives of the covariance with respect to $\fnl$, we use the fact that
the derivative of the bias with respect to $\fnl$ is analytic so that
\begin{eqnarray}
\frac{d\xi^\alpha_{ij}}{d\fnl} &\equiv& 
\int \frac{d^3 k}{ (2\pi)^3} |W({\bf k})|^2 \cos(k_x \Delta x_{ij})\cos(k_y \Delta y_{ij}) \nonumber  \\
&&\times \cos(k_z \Delta z_{ij}) \frac{d(b_{i\alpha}b_{j\alpha})}{d\fnl} P(k;z).
\label{eq:dsij}
\end{eqnarray}
In calculating $dS_{ij}/d\fnl$, we only keep the dominant term, which is the
one with derivative with respect to $\xi_{ij}^{\alpha}$.  That is, we assume
that
\begin{eqnarray}
\frac{dS^{\alpha}_{ij}}{d\fnl}\simeq
\bar m_{i\alpha} \bar m_{j\alpha}\frac{d\xi^{\alpha}_{ij}}{d\fnl}.
\end{eqnarray}
\noindent The terms we ignore correspond to the sensitivity of cluster counts
to non-Gaussianity, and they would only {\it enhance} the impact of $\fnl$,
though slightly, as will be shown in the following sections.  In a real survey
one actually has to calculate the covariance at non-zero values of $\fnl$ for
which our approach of evaluating the derivative analytically at $\fnl=0$ would
be insufficiently general.  For this sensitivity study, however, the analytic
derivative is perfectly acceptable.  We examine the sensitivity to the
constraints around different fiducial values of $\fnl$ in Sec. \ref{sec:disc}.

\begin{figure*}[!t]
\includegraphics[scale=0.40]{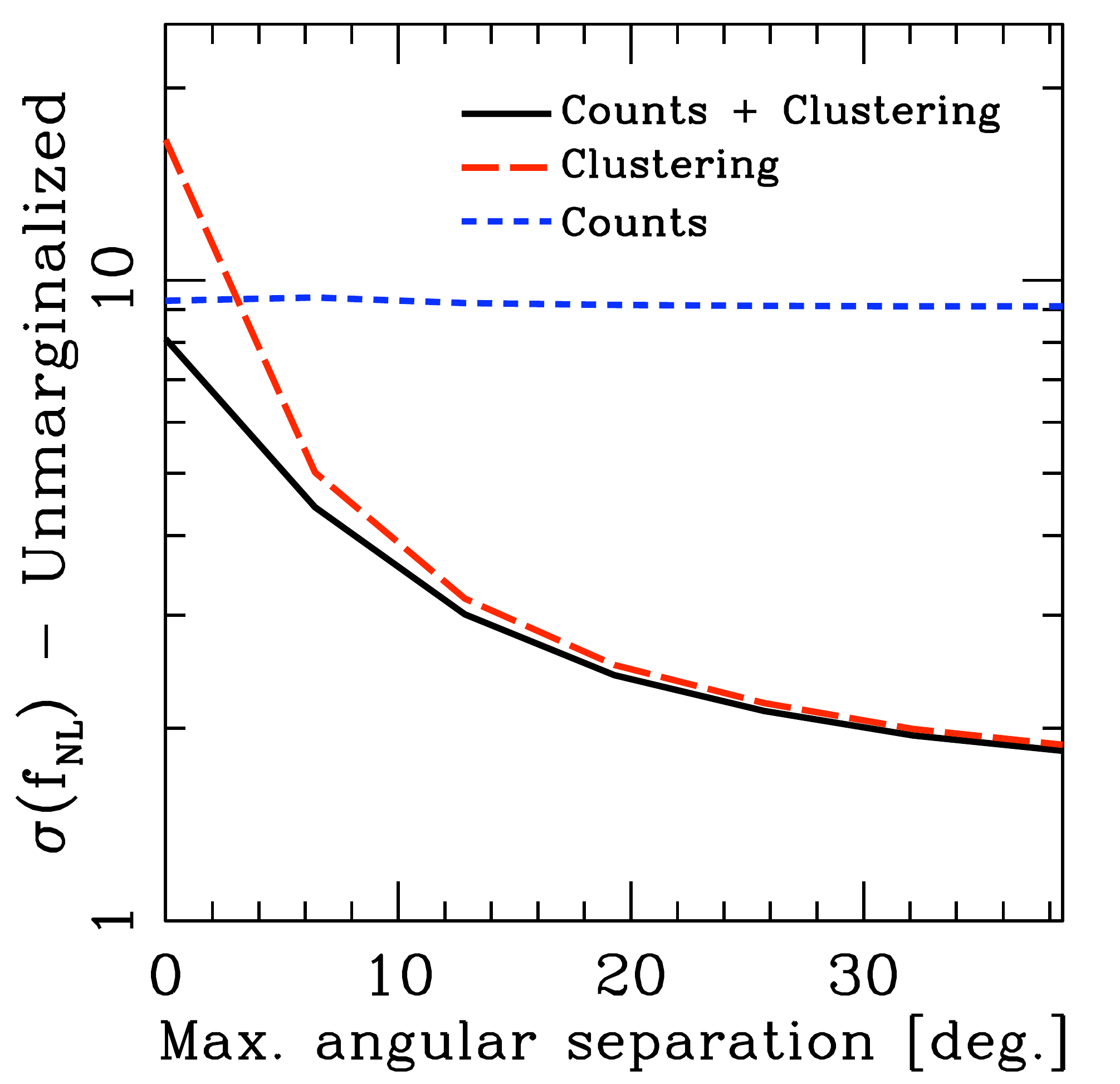}\hspace{0.5cm}
\includegraphics[scale=0.40]{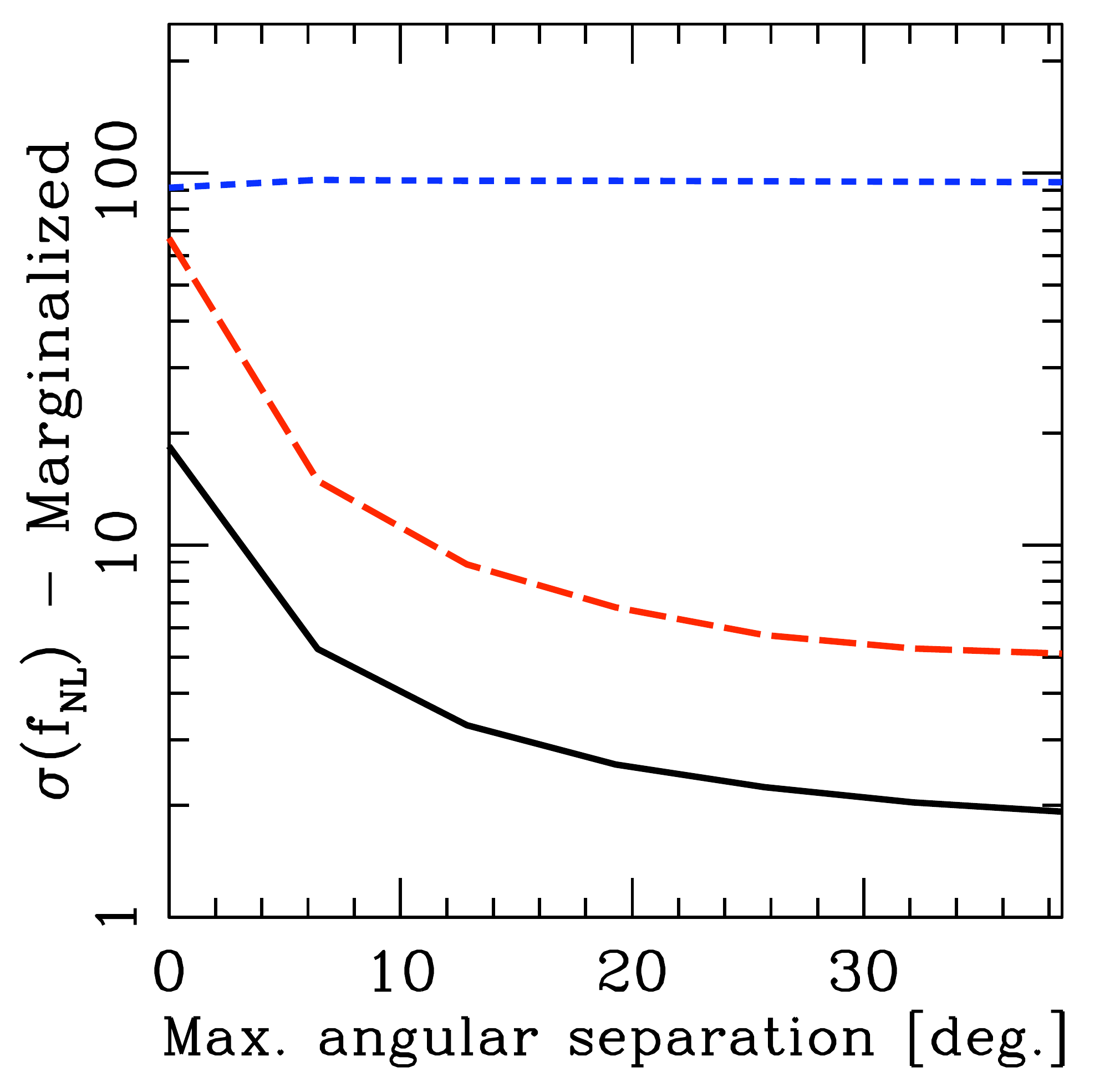}
\caption{ $1-\sigma$ uncertainties in the parameter $\fnl$ as a function of the maximum angular
  separation between pixel centroids in the covariance matrix. The {\it left}
  panel shows the unmarginalized constraints while the {\it right} panel shows
  marginalized constraints assuming Planck priors and fixed halo-bias and
  observable-mass nuisance parameters. Zero separation indicates the case of
  pure variances (as considered by \citet{Oguri09}). The maximum angular
  separation between pixels for a $5,000$ sq. deg. survey divided into 41.3 sq. deg
  pixels is about 90 degrees (or $10\sqrt{2}$ pixel widths).  This case would
  correspond to taking the full covariance into account for the calculation of
  $\fnl$, but disregarding the covariance between different redshift bins.  
  The {\it blue short dashed} line corresponds to constraints derived
  using only cluster counts.  The {\it red dashed} line shows the constraints
  when only the clustering of clusters is used, and the {\it solid black} line
  shows the combined constraints from counts and clustering.  }
\label{fig:nsep}
\end{figure*}

\section{Results}\label{sec:results}

Our results are presented as follows.  First we discuss the sensitivity of
cluster counts and clustering of counts to $\fnl$, and examine unmarginalized
constraints on $\fnl$.  Second, we examine the degeneracies with
cosmological parameters and nuisance parameters  due
to modeling uncertainties in the observable-mass relation and in the halo
bias.  Third and last, we look at the impact of photometric redshift uncertainties.

\subsection{Sensitivity of cluster covariance}\label{sec:sens}

The effect of non-Gaussianity on clustering is a combination of several
effects, which can be identified from Eq.~(\ref{eq:Sij}).  The dominant effect
is due to the explicit modification of the halo bias (Eq.~(\ref{eq:bias}))
which affects $\xi^{\alpha}_{ij}$ (cf.\ Eq.~(\ref{eq:Sij_bare})) In addition,
non-Gaussianity affects the mass function, which affects the mean cluster
counts (cf.\ Eqs.~\ref{eqn:nofz} and \ref{eqn:numwin}), and the average
cluster linear bias (cf.\ Eq.~(\ref{eqn:linbias})).  The left panel of
Fig.~\ref{fig:Sij} shows the dependence of the different terms that make up
the clustering covariance $S_{ij}$, as a function of $\fnl$.  For this
sensitivity plot, we assume a 40 sq.\ deg.\ pixel with redshift thickness
$\Delta z=0.2$ centered around $z=0.5$ and a mass-threshold $\Mth=10^{13.7}
h^{-1} \Msun$, and show only the diagonal elements $i=j$ for clarity.  The
relation between the functions plotted in this figure is $S_{ij}={\bar
  m}^2\xi^{\alpha}_{ij}$.  It is apparent from the figure that
$\xi^{\alpha}_{ij}$ encodes most of the dependence of the clustering signal on
$\fnl$, and that the clustering covariance ($S_{ij}$, or $\xi^{\alpha}_{ij}$)
is much more sensitive to $\fnl$ than the mean counts $\bar m$.  As mentioned
previously, we neglect the implicit mass function dependence of $\fnl$ when
calculating the covariance.  Including it would only enhance the
impact of $\fnl$, albeit slightly.

In the right panel of Fig.~\ref{fig:Sij} we plot the absolute value of the
clustering covariance as a function of angular separation between the
centroids of two pixels.  For reference, at $z=0.5$, a one-degree separation
corresponds to about $23.4 h^{-1}\Mpc$.  For $\fnl=0$, the clustering
covariance is large and positive at small separations, but becomes negative at
intermediate pixel separations ($\sim 6$ deg or $\sim 150 \hmpc$ at $z=0.5$);
this behavior corresponds to a similar behavior of the two-point correlation
function $\xi(r)$ (see e.g.\ Ref.~\cite{Estrada_Frieman2009}).  The effect of
nonzero $\fnl$ depends on its sign as well as on the scale.  For positive
$\fnl$, the covariance increases monotonically with $\fnl$ roughly up to the
scale of the survey.  Beyond that scale ($\sim 60\degr$ in our example), the
covariance reverses its trend with $\fnl$ and becomes negative due to the
integral constraint imposed by the regularization.  For negative $\fnl$, the
dependence of the covariance $S_{ij}$ on $\fnl$ is more complicated because
the total bias becomes negative at large enough scales; thus, for $\fnl<0$ the
covariance depends monotonically on $|\fnl|$ only on scales ($\lesssim 7\degr$
in the right panel of Fig.~\ref{fig:Sij}) for which the bias correction --
second term in Eq.~(\ref{eq:bias}) -- is subdominant.  Note that
Fig.~\ref{fig:Sij} hides the fact that the number of pixels at a given
separation increases with separation: the number of off-diagonal elements in
the covariance is much bigger than the number of diagonal elements, and this
gives a ``'geometric boost'' to the covariance.

\subsection{Unmarginalized constraints from clustering and counts}\label{sec:unmarg}

Both panels of Fig.~\ref{fig:nsep} show $\fnl$ constraints as a function of
the maximum pixel separation allowed in the covariance (cf. Eq. \ref{eq:Sij})
used to generate the Fisher matrix constraints (cf. Eq. \ref{eq:fisher}).

\begin{table}[!t]
\begin{center}
\begin{tabular}{c|c|cc|c}
\hline\hline         \multicolumn{4}{c}{\rule[-2mm]{0mm}{6mm}Unmarginalized error $\sigma(\fnl)$}\\
\hline\hline          Assumption & \rule[-2mm]{0mm}{6mm} Number
& \rule[-2mm]{0mm}{6mm} Counts 
& \rule[-2mm]{0mm}{6mm} Covariance    
& \rule[-2mm]{0mm}{6mm} {\bf Both} \\\hline\hline
\rule[-2mm]{0mm}{6mm}Fiducial      &$1.7\times10^5$  & 9.1 & 1.8   & {\bf 1.7} \\
\rule[-2mm]{0mm}{6mm}12.5 $\sqdeg$ pix  & $1.7\times10^5$ & 9.2 & 1.1   & {\bf 1.1} \\
\rule[-2mm]{0mm}{6mm}$\zmax=0.8$   & $1.3\times10^5$ & 13 & 2.3   &{\bf 2.2} \\
\rule[-2mm]{0mm}{6mm}$\zmax=1.4$   & $2.4\times10^5$ & 6.0 & 1.4   &{\bf 1.4} \\
\rule[-2mm]{0mm}{6mm}$\Mth=10^{13.5}$ &$3.6\times10^5$ & 8.3 & 1.4   &{\bf 1.4} \\
\rule[-2mm]{0mm}{6mm}$\Mth=10^{13.9}$&$7.7\times10^4$  & 10 & 2.3  &{\bf 2.3} \\\hline \hline
\end{tabular}
\caption{Unmarginalized constraints on $\fnl$. The fiducial case assumes no
  nuisance parameters, 5 bins in mass and redshift each, and other assumptions
  as in the text. Variations in the assumptions are shown in the first column,
  followed by the total number of clusters in the $5,000\sqdeg$ survey we
  assumed, while cluster counts, covariance, and combined projected 1-$\sigma$
  constraints on $\fnl$ are given in the following three columns.  }
\label{tab:unmarg}
\end{center}
\end{table}

In the left panel of Fig.~\ref{fig:nsep} we see that the cluster counts yield
better unmarginalized constraints than the 
{\it variance} of cluster counts alone; however, once the covariances
(i.e.\ off-diagonal terms of the signal matrix $S_{ij}$) are included, the
clustering information rapidly beats that from the counts.  In Table
\ref{tab:unmarg} we show the {\it unmarginalized} $\fnl$ constraints for a
variety of survey expectations.  Changes in the constraints improve in the
direction expected: the lower the mass-threshold and the higher the maximum
redshift, the better.  This Table also shows that decreasing the angular area
of the pixels to $12.5\sqdeg$ results in substantial ($O(50\%)$) improvements.
The improvement with decreasing pixel size, for $\fnl$ constraints, does not
happen if we consider only the variance in counts.  For other parameters, that
are sensitive to small scale information, such as $\DE$ and $w$, the smaller
pixels do translate into better constraints even if only the sample variance
is used.  Further refinements of the pixelization leads to improvement up to
the regime of shot-noise domination, (which occurs for pixels of $\sim
0.1-1\sqdeg$).  Unmarginalized constraints are of order $10^{-1}$ in this
regime, though observational systematics are likely to dominate 
statistical errors of this size.

\subsection{Degeneracies with cosmological and nuisance parameters.}\label{sec:marg}

In the right panel of Fig.~\ref{fig:nsep} we show the {\it marginalized}
constraints on $\fnl$ assuming Planck priors and fixed nuisance parameters
(both halo bias and mass-observable).  We see that the change in
the constraints from combined counts\footnote{The slight degradation in $\fnl$
  constraints from counts seen in the right panel is real, and is due to
  adding the (positive) covariance matrix elements to the counts noise; see
  the first term on the RHS of Eq.~(\ref{eq:fisher}).  Using the full covariance
  therefore yields very slightly worse constraints.} and clustering is even
more remarkable than the unmarginalized constraints shown in the right panel.
The full clustering covariance yields about one order of magnitude better
constraints than if only the variance is used.  As we shall see, this
fractional improvement remains even when we include nuisance parameters.

\begin{table*}[!t]
\begin{center}
\begin{tabular}{cccccccccccccc}
\hline\hline \multicolumn{14}{c}{
\rule[-2mm]{0mm}{6mm}{\bf Marginalized errors - Variance only}}\\
\hline \hline 
\multicolumn{2}{c}{Nuisance parameters}& & \multicolumn{3}{c}{Counts}& &\multicolumn{3}{c}{Variance}& &\multicolumn{3}{c}{Counts+Variance} \\ 
Halo bias & $\Mobs$  & & \rule[-2mm]{0mm}{6mm} $\sigma(\DE)$ & \rule[-2mm]{0mm}{6mm} $\sigma(w)$ & \rule[-2mm]{0mm}{6mm} {\bf ${\mathbf \sigma(\fnl)}$ }&&  \rule[-2mm]{0mm}{6mm} $\sigma(\DE)$ & \rule[-2mm]{0mm}{6mm} $\sigma(w)$ & \rule[-2mm]{0mm}{6mm} {\bf ${\mathbf \sigma(\fnl)}$ }& & \rule[-2mm]{0mm}{6mm} $\sigma(\DE)$ & \rule[-2mm]{0mm}{6mm} $\sigma(w)$ & \rule[-2mm]{0mm}{6mm} {\bf ${\mathbf \sigma(\fnl)}$ } \\
\hline
\rule[-2mm]{0mm}{6mm}Marginalized & Marginalized & & $\infty$ & $\infty$   & {$ \bf \infty$ } & & $\infty$ & $\infty$   & { $\bf \infty$} & & 0.075 & 0.25   & {\bf 55}\\ 
\rule[-2mm]{0mm}{6mm}Known & Marginalized & & 0.095 & 0.32  & { $\bf 3.4 \times 10^{3}$}  && $\infty$ & $\infty$   & {$\bf \infty$} && 0.061 & 0.21   & {\bf 27}  \\
\rule[-2mm]{0mm}{6mm}Marginalized & Known&  & $\infty$ & $\infty$   & {$ \bf \infty$}  && 0.077 & 0.26   & {\bf 98}& & 0.0037 & 0.016   & {\bf 44} \\
\rule[-2mm]{0mm}{6mm}Known & Known &  & 0.0046 & 0.021   & {\bf 91}  && 0.053 & 0.18   & {\bf 67} & & 0.0035 & 0.014   & {\bf 19}\\\hline \hline
\end{tabular}
\caption{Marginalized constraints on $\fnl$ and dark energy with cluster
  counts, variance of the counts, and the two combined. The fiducial case
  assumes 5 bins in mass and redshift each with a mass threshold
  $\Mth=10^{13.7}$, maximum redshift $\zmax=1.0$, and other assumptions as in
  the text. Assumptions about the nuisance parameters are varied, and are
  shown in the first two columns.  Entries with $\infty$ indicate
  that the method was unable to constrain the parameters. }
\label{tab:margvarbig}
\end{center}
\end{table*}

\begin{table*}[!t]
\begin{center}
\begin{tabular}{cccccccccccccc}
\hline\hline \multicolumn{14}{c}{
\rule[-2mm]{0mm}{6mm}{\bf Marginalized errors - Full Covariance}}\\
\hline \hline 
\multicolumn{2}{c}{Nuisance parameters} && \multicolumn{3}{c}{Counts}&& \multicolumn{3}{c}{Covariance}&& \multicolumn{3}{c}{Counts+Covariance} \\ 
Halo bias & $\Mobs$  & & \rule[-2mm]{0mm}{6mm} $\sigma(\DE)$ & \rule[-2mm]{0mm}{6mm} $\sigma(w)$ & \rule[-2mm]{0mm}{6mm} {\bf ${\mathbf \sigma(\fnl)}$ }& & \rule[-2mm]{0mm}{6mm} $\sigma(\DE)$ & \rule[-2mm]{0mm}{6mm} $\sigma(w)$ & \rule[-2mm]{0mm}{6mm} {\bf ${\mathbf \sigma(\fnl)}$ }& & \rule[-2mm]{0mm}{6mm} $\sigma(\DE)$ & \rule[-2mm]{0mm}{6mm} $\sigma(w)$ & \rule[-2mm]{0mm}{6mm} {\bf ${\mathbf \sigma(\fnl)}$ } \\
\hline
\rule[-2mm]{0mm}{6mm}Marginalized & Marginalized & & $\infty$ & $\infty$   & {\bf $\infty$ } & & $\infty$ & $\infty$   & {\bf $\infty$} & &0.069 & 0.23   & {\bf 6.0 } \\ 
\rule[-2mm]{0mm}{6mm}Known & Marginalized & & 0.097 & 0.33  & {$\bf 2.1 \times 10^{3}$ }  && 0.13 & 0.43  & {\bf 12}& & 0.065 & 0.22   & {\bf 5.4}  \\
\rule[-2mm]{0mm}{6mm}Marginalized & Known & & $\infty$ & $\infty$   & {\bf $\infty$} & & 0.099  & 0.34  & {\bf 7.0 } && 0.0036 & 0.014  & {\bf 3.8} \\
\rule[-2mm]{0mm}{6mm}Known & Known  & & 0.0051 & 0.023   & {\bf 94 } & & 0.042 & 0.13  & {\bf 5.1 } & & 0.0036 & 0.014  & {\bf 1.8} \\\hline \hline
\end{tabular}
\caption{Marginalized constraints on $\fnl$ and dark energy with cluster
  counts, covariance of the counts, and the two combined. The fiducial case
  assumes 5 bins in mass and redshift each with a mass threshold
  $\Mth=10^{13.7}$, maximum redshift $\zmax=1.0$, and other assumptions as in
  the text.  Assumptions about the nuisance parameters are varied, and are
  shown in the first two columns.  Entries with $\infty$ indicate that the
  method was unable to constrain the parameters.}
\label{tab:margcovarbig}
\end{center}
\end{table*}

Tables \ref{tab:margvarbig} and \ref{tab:margcovarbig} show $\fnl$ constraints
using the 
variance of cluster counts,  and the full 
covariance, respectively.  The results assumed Planck priors on the
cosmological parameters, 10 nuisance parameters describing the mass-observable
relation and 3 nuisance parameters describing uncertainties in the Gaussian
halo bias.

Comparing the last columns of Tables \ref{tab:margvarbig} and
\ref{tab:margcovarbig}, we see that the counts+covariance combination yields
about an order of magnitude improvement over simply using counts+variance.
For the counts+variance, the uncertainties in the halo bias parameters are the
main source of degradation to $\fnl$ constraints.  Without the information
from large separations provided by the full covariance, the Fisher matrix
cannot disentangle the effects due to the Gaussian bias from the $\fnl$
contribution.  When the full covariance is used (cf. Table
\ref{tab:margcovarbig}), the errors in the mass-observable relation are the
dominant source of degradation.  Marginalizing over all nuisance
parameters, assuming flat priors, yields a degradation of $\sim 3$ in
$\sigma(\fnl)$.  This is not large, considering we added 13 nuisance
parameters, but not negligible either.  
Even modest prior information
can improve the marginalized constraints significantly.

There are two principal reasons for the strong improvement of errors when the
covariance is added:

\begin{enumerate}
\item  The strong scale dependence of the bias as a function 
implies that most signal comes from the covariances, since the covariances have
longer lever arms in $k$ than the variance alone (and are much more sensitive
than counts which only depend on non-Gaussianity via the mass function);
\item The signature of $\fnl$ in the covariance is unique, as no other
cosmological parameter leads to a similar effect --- therefore, the degeneracy
with other cosmological parameters is very small, as first noted by
\cite{Dalal}.  
\end{enumerate}
Comparing the $\fnl$ constraints for the full covariance for fixed nuisance
parameters (Table \ref{tab:margcovarbig}) to the unmarginalized constraints
(Table \ref{tab:unmarg}), we see that degeneracies with cosmological
parameters only result in a small degradation of $\fnl$ constraints (from 1.7
to 1.8).

Tables \ref{tab:margvarbig} and \ref{tab:margcovarbig} also show the
constraints obtained using counts alone, or (co)variance by itself.  The
information about $\fnl$ from the {\it counts} is very degenerate with the
cosmological and nuisance parameters.  The ``$\infty$'' symbols indicate that
the Fisher matrix could not be inverted, i.e., that particular technique was
unable to simultaneously constrain all of the parameters.  From the last row
of both tables, we see that cluster counts are effective at constraining the
cosmological parameters and mass-observable relation (from the mass binning)
whereas the (co)variance constrains mainly the nuisance parameters and
$\fnl$.

Marginalization degrades the counts + covariance $\fnl$ constraints roughly
independently of the different survey assumptions, so one can use Table
\ref{tab:unmarg} to infer marginalized constraints.  For example, from Table
\ref{tab:unmarg}, we see that using $12.5\sqdeg$ pixels yields about 60\%
better constraints.  The full marginalized constraints are also improved by a
similar factors so that, for example, $\sigma({\fnl}) \sim 3.9$ for
$12.5\sqdeg$ marginalized over the 13 nuisance parameters (compared to
$\sigma({\fnl}) =6.0$ for $40\sqdeg$ pixels).

\subsection{Photometric redshift errors}\label{sec:photoz}

To study the effects of photometric redshift errors, we add 10 nuisance
parameters to the analysis, namely two parameters --- one each describing the
photo-z scatter and bias --- in each of the five redshift bins.  The results
are summarized in Table \ref{tab:photoz}.

If either the halo bias or the mass-observable nuisance parameters are fixed,
then the degradation from the inclusion of photo-z's is not very damaging. In
other words, the additional correlations between either photo-z and halo bias
parameters, or between photo-z and mass-observable parameters, do not cause
substantial additional degradation to $\fnl$ constraints (relative to the case
where only the photo-z parameters are unknown).

However when all 23 nuisance parameters (10 for the photo-z's, 10 for the
mass-observable relation, and 3 for halo bias) are left free, one cannot
simultaneously constrain dark energy and $\fnl$, and the constraints on both
drastically degrade.  
We traced the biggest source of degradation to the redshift evolution
parameters in the mass-observable relation and to the photo-z bias nuisance
parameters.  Simply adding a 33\% prior to the one parameter describing the
evolution of the bias in $P(\Mobs|M)$ (parameter $a_1$ in
Eq.~(\ref{eqn:mbiasdef})) was enough to reclaim respectable accuracy, with
$\sigma(\fnl)=18.8$ (see the bottom row of Table \ref{tab:photoz}).
Alternatively, if the bias in each photo-z bin is known to the absolute
accuracy of $0.01$ with all other parameters free, then $\sigma(\fnl)=7.0$, which
is just $\sim 15\%$ worse than when photo-z parameters are
fixed\footnote{Unlike $\fnl$, the dark energy constraints are sensitive to
  both bias and scatter of the photo-z's.  For a prior uncertainty in the
  photo-z bias of $0.01$ per bin, the photo-z scatter needs to be known to
  $0.025$ {\it per bin} to achieve small ($\lesssim 15\%$)
degradation in $\sigma(\DE)$ and $\sigma(w)$ relative to the case of perfectly
known photo-z errors. }. For a survey such as the DES, these requirements
should be relatively easy to satisfy, given that spectroscopic samples of
$10^4$-$10^5$ galaxies will be available to calibrate the photometric redshift errors
(see e.g.\ Eqs.~(19) and (20) in \citet{Hearin}).

\begin{table}[!t]
\begin{center}
\begin{tabular}{ccccc}
\hline\hline \multicolumn{5}{c}{\rule[-2mm]{0mm}{6mm}The effects of photo-z uncertainties}\\
\hline \hline \multicolumn{2}{c}{Nuisance parameters }& \multicolumn{3}{c}{} \\ 
Halo bias & $\Mobs$  &  \rule[-2mm]{0mm}{6mm} $\sigma(\DE)$ & \rule[-2mm]{0mm}{6mm} $\sigma(w)$ & \rule[-2mm]{0mm}{6mm} {\bf $\sigma(\fnl)$ } \\
\hline
\rule[-2mm]{0mm}{6mm}Known & Known  & 0.016 & 0.041  & {\bf 6.5 } \\
\rule[-2mm]{0mm}{6mm}Marginalized & Known  & 0.021 & 0.053  & {\bf 6.7 } \\
\rule[-2mm]{0mm}{6mm}Known & Marginalized  &0.11  & 0.36   & {\bf 9.4 } \\
\rule[-2mm]{0mm}{6mm}Marginalized$^a$ & Marginalized$^a$  &0.23$^a$  & 0.77$^a$  & {\bf 19$^a$} \\\hline \hline
\end{tabular}
\caption{ Effect of photometric redshift uncertainties on the marginalized
  constraints on $\fnl$. The fiducial case assumes 5 bins in mass and redshift
  each with a mass-threshold $\Mth=10^{13.7}$ and maximum redshift
  $\zmax=1.0$, and other assumptions as in the text. Variations are in the
  first two columns, while cluster, covariance, and combined projected
  1-$\sigma$ constraints on $\fnl$ are given in the following three columns.\\
  In the bottom row, superscript $^a$ signals that a Fisher matrix prior of
  $F_{a_1,a_1}=10$ is added to the nuisance parameter $a_1$ defined in
  Eq.~(\ref{eqn:mbiasdef}), which describes the redshift evolution of the bias
  in the mass-observable relation.  }
\label{tab:photoz}
\end{center}
\end{table}

\section{Discussion}\label{sec:disc}

\subsection{Choice of the fiducial model}\label{sec:disc_fiducial}

In our fiducial approach we estimated errors in $\fnl$ around
$\fnl=0$. However, it is a slightly different matter to estimate the {\it
  detectability} of non-Gaussianity, which requires estimating the
signal-to-noise at which a {\it non-zero} fiducial value of $\fnl$ can be
differentiated from zero\footnote{Arguably the best approach might be to use
  the Bayesian model selection techniques and, for a range of $\fnl$ values,
  test if the hypothesis $\fnl=0$ can be rejected. We do not pursue such an
  approach in this paper.}. The detectability is independent of the fiducial
value if the observable quantity is linear in the parameter(s); this is
clearly not the case here since the clustering signal is a quadratic function
of the bias, which itself depends linearly on $\fnl$.

\begin{figure*}[!t]
\includegraphics[scale=0.40]{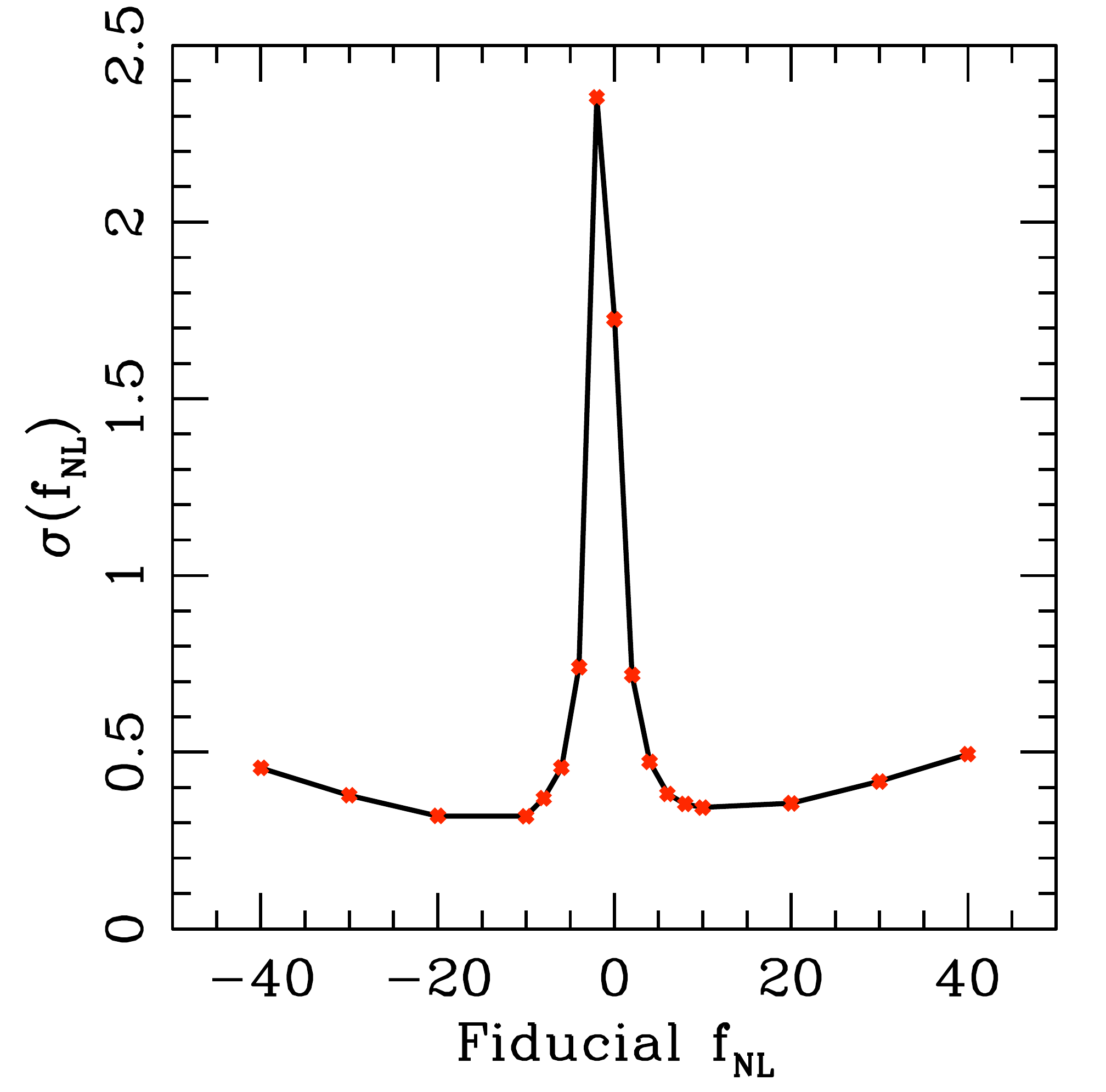}\hspace{0.5cm}
\includegraphics[scale=0.40]{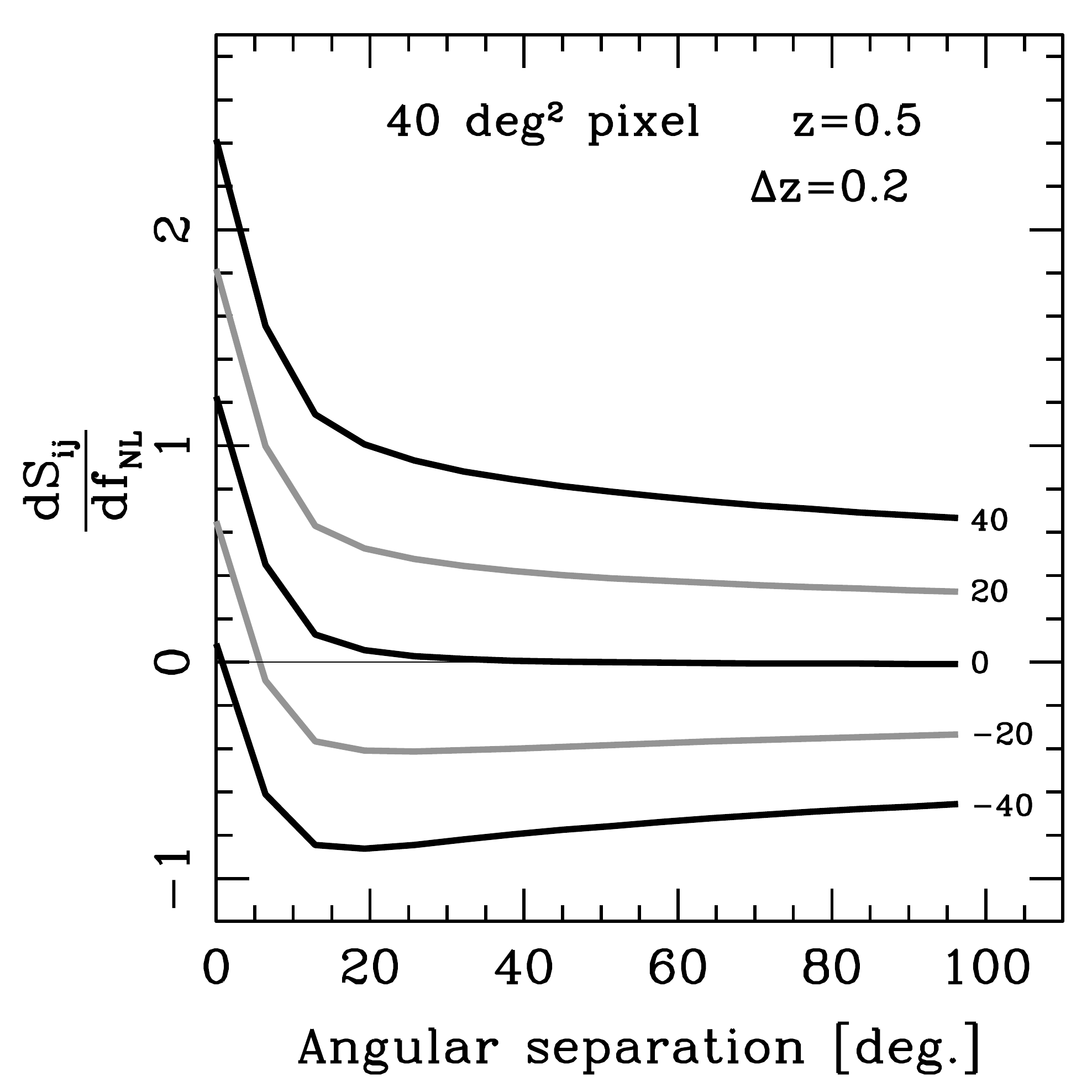}
\caption{Left panel: Unmarginalized $1-\sigma$ constraints on $\fnl$ as a function of the
  fiducial value of this parameter, assuming five redshift and five
    mass bins. The ``witch's hat'' shape can be explained from the
  competition between the derivative of the covariance with respect to $\fnl$,
  and the total covariance at the fiducial $\fnl$; see text. Right panel:
  Derivative of the signal matrix elements $S_{ij}$ with respect to $\fnl$ as
  a function of angular separation between pixels $i$ and $j$, for $\fnl=-40,
  -20, 0, 20$, and $40$.  Recall that, at $z=0.5$, a separation of 1 degree
  corresponds to about $23h^{-1}$Mpc.}
\label{fig:varfid}
\end{figure*}

Fig.~\ref{fig:varfid} shows the fiducial unmarginalized 
constraints on $\fnl$ as a function of its fiducial value.  Unlike in the
results shown previously, here we calculate all elements of the covariance
matrix and its derivative with respect to $\fnl$ (which is why the constraints
for $\fnl=0$ shown in the plot are slightly better than what is shown in Table
\ref{tab:unmarg}).  The figure shows tightest constraints for $|\fnl|\simeq
10$ --- more than 4 times stronger than those for our fiducial assumption of
$\fnl=0$. The ``witch's hat'' shape shown in Fig.~\ref{fig:varfid} can be
understood by examining the second term on the RHS of Eq.~(\ref{eq:fisher})
that contains the Fisher information from the covariance of cluster counts.
The $\fnl$ constraints are set by the competition between the signal,
represented by the derivative of the covariance with respect to $\fnl$, ${\bf
  S}_{,\mu}$, and the noise, given by the total covariance, ${\bf C}$. These
two quantities vary with $\fnl$ at different rates; the total covariance
depends (roughly) quadratically on $\fnl$ whereas ${\bf S}_{,\mu}$ only has a
linear dependence.  In addition, the matrix elements of ${\bf S}_{,\mu}$ and
${\bf C}$ have different sensitivity to $\fnl$ at each angular separation, and
it is the relative importance of the off-diagonal matrix elements relative to
the diagonal elements that sets the shape of the curve in
Fig.~\ref{fig:varfid}.

For very small values of $|\fnl|$ ($\ll 10$), the off-diagonal elements of the
covariance are very small, and hence do not contribute much to the signal, or
to ${\bf C}$.  This can be seen in the $\fnl= 0$ curves in the right panel of
Fig.~\ref{fig:Sij} and in the right panel of Fig.~\ref{fig:varfid}.  Note that
the plots hide the fact that the number of pixels at a given separation 
increases with separation: the number of off-diagonal elements in the
covariance is much bigger than the number of diagonal elements, and this gives a
``'geometric boost'' to the covariance.

For large values of $|\fnl|$ ($\gg 10$), the off-diagonal
elements of the covariance matrix 
can be significant relative to the diagonal elements
(see the $\fnl=\pm 100$ curves in the right panel of Fig.~\ref{fig:Sij}).
Therefore, the constraints on $\fnl$ now worsen with the increasing value of
$|\fnl|$, albeit slowly.

Finally, in the intermediate range of $|\fnl| \sim 10$, the off-diagonal
elements of ${\bf C}$ are small relative to the diagonal and near-diagonal
elements. For example, the right panel of Fig.~\ref{fig:Sij} shows that, for
$\fnl=20$, the far-separation covariances are much smaller than the variances.
However the derivatives of the sample covariance, $d{\bf S}/d\fnl$, are only
moderately smaller for the off-diagonal pixels than for the diagonal ones
(e.g.\ a factor of $\sim 4$ for $\fnl=20$; see the right panel of
Fig.~\ref{fig:varfid}). Therefore, it is at these intermediate values of
$|\fnl|\sim 10$ that we find the best signal-to-noise, and best constraints on
$\fnl$.

In summary, the dependence on the fiducial value of $\fnl$ can be understood
rather simply. For small $\fnl$, the large-scale covariances do not add much
signal.  For large $\fnl$ the covariances add too much noise. At intermediate
$\fnl$, the signal-to-noise relation is ``just right''.  We caution that the
shape of the curve in Fig.~\ref{fig:varfid} depends on the volume (and
geometry) of the survey as well as in the number density of sources.  The
width of the pixels affect the width of the central part of the ``hat''
slightly.  Smaller bins tend to shift the minima to smaller values of $|\fnl|$.
We conclude that the power of a DES-like cluster surveys to rule out the
Gaussian hypothesis may be even greater than indicated in Tables in this paper, 
since the error at $\fnl\neq 0$ nearly always smaller than that for $\fnl=0$. 
This is another exciting development, but warrants further investigation, 
and in particular a more detailed study of the dependencies on the overall 
survey volume and selection.
In this initial study we simply adopt the conservative errors, and show the
$\fnl=0$ results everywhere except in Fig.~\ref{fig:varfid}.

\subsection{Clusters vs.\ galaxies}\label{sec:clus_gal}

It is useful to compare cluster constraints obtained here with the expected
constraints from a similar, DES-type, galaxy survey. Forecasts of constraints
on primordial non-Gaussianity from galaxy clustering were studied recently
\cite{Dalal,Carbone,Sartoris} using the Fisher matrix and a simple,
Feldman-Kaiser-Peacock (FKP \cite{FKP}) estimator that counts modes of $P(k)$
and combines them with the survey volume and its galaxy
density. Perhaps counterintuitively, our constraints are a factor of a few
better than those from galaxies estimated previously. We now explain the
origin of this apparent discrepancy.

Both clusters and galaxies probe the power spectrum of dark matter halos (and
thus the halo bias). However, there are some important differences
\begin{itemize}
\item Clusters additionally probe the mass function, which
determines the counts, and also weakly affects the bias $b_0(M, z)$; see
Eqs.~(\ref{eq:sigma}) and (\ref{eqn:linbias});
\item The number density of galaxies may be significantly higher, depending on
  how they and the clusters are selected. However, as mentioned in
  Sec.~\ref{sec:calc}, the larger size of galaxy samples may not bring much
  additional information, since the constraints on $\fnl$ benefit from very
  large-scale halo separations, and not from intra-halo correlations; 
\item Clusters reside in more massive halos than galaxies, and thus have a
  higher bias. The higher the bias, the stronger is the correlation 
  (cf. Eq. \ref{eq:Sij_bare});
\item With regards to systematics, clusters can naturally be binned by the
  mass-observable, which helps break degeneracies with nuisance
  parameters. This allows utilization of the cross-correlation between
  different mass bins to reduce the impact of sample variance
  (e.g. \cite{Seljak09,McDonald_Seljak09}), which we do not exploit
  in this paper. 
\item Large spectroscopic samples of galaxies are expected in the near future,
  whereas clusters will rely on photometric redshifts; therefore, galaxy 
  redshifts are likely to be more accurate than cluster redshifts;
\end{itemize}
Given all these differences, it is difficult to predict whether clusters or
galaxies will give a stronger constraints on primordial non-Gaussianity
without a direct calculation.  We have verified that the FKP estimator of
galaxy constraints on $\fnl$ indeed gives a weaker result, and is in rough
agreement with previous estimates in \cite{Dalal,Carbone,Carbone2}.

However, as discussed in \citet{Tegmark_etal}, the FKP estimator is only optimal and lossless on
scales much smaller than the linear size of the survey.
Since good constraints on $\fnl$ benefit from precisely the large-wavelength modes, it is not
surprising that the FKP estimator for galaxies indicates worse constraints
than our pixel-based estimator for clusters. We have additionally verified
that constraints on the constant part of the bias, $b_0$ (see
Eq.~(\ref{eq:bias})), or the dark energy equation of state $w$, which do not 
benefit as much from large-wavelength modes, are comparable when estimated from 
the pixel-based formalism (from this paper) and the FKP approach assuming 
the same survey volume and number density of objects.

 \subsection{Comparison to previous work}

 Numerous papers have studied the power of cluster counts alone to probe
 primordial non-Gaussianity (e.g.\
 \cite{Verde_CMBLSS,MVJ,Verde_tests,Sefusatti,Dalal}). To the extent that such
 constraints are generally weak due to degeneracies, and strongly depend on
 the priors and nuisance parameters varied, our results (see the ``counts''
 columns in Table \ref{tab:margcovarbig}) are in broad agreement with these
 studies.

 A more interesting comparison can be made with the recent work of
 \citet{Oguri09} who studied the counts+variance case of clusters,
 corresponding to results in our Table \ref{tab:margvarbig}. The main
 difference between the two studies is that we additionally considered the
 covariance of cluster counts, and found that it leads to a huge further
 improvement in the constraints.  However, even for the counts+variance case
 only, our results differ substantially, and we forecast a much {\it weaker}
 constraint on non-Gaussianity than Ref.~\cite{Oguri09}. For example, we get
 $\sigma(\fnl)\sim 20$-$30$ compared to $\sigma(\fnl)\sim 8$ in
 Ref.~\cite{Oguri09} in the most fair comparison with their DES survey case
 and our assumptions with either no nuisance parameters or full
 mass-observable nuisance parameters\footnote{Ref.~\cite{Oguri09} assumes only two
   mass-observable nuisance parameters.}.  These discrepancies could
 probably be explained by a number of other differences in the analyses: mass
 functions (Ref.~\cite{Oguri09} uses the \citet{LoVerde} mass function with
 analytic fit for skewness, while we use Dalal et al.\ mass function from
 Eqs.~(\ref{eq:mean_Mf_def})-(\ref{eq:mf_conv})); cosmological parameter
 priors (Ref.~\cite{Oguri09} uses the diagonal priors on some parameters while
 we use the full, off-diagonal Planck prior Fisher matrix), etc.  We have not
 attempted to reproduce results from Ref.~\cite{Oguri09} using the assumptions
made in that paper. 

\subsection{Issues for future study}

There are a number of effects that remain to be studied in detail, but are
beyond the scope of this preliminary analysis. We now list them here:

\begin{itemize}
\item {\it Fisher matrix approximation:} in this paper we have assumed the fiducial
  value of $\fnl=0$ and calculated the errors on $\fnl$ by taking the
  derivatives of observables with respect to this parameter. This ``Fisher
  error'' will be a good approximation to the true error if the error itself
  is small. Therefore, at least in the cases where the $\fnl$ error is tight,
  we expect the Fisher approximation is a good one, though this should
  eventually be checked with Markov chain Monte Carlo methods.

\item {\it Calculational issues:} The computation of the cluster covariance is
  time consuming, particularly for small but non-zero values of $\fnl$. In this
  work we have largely avoided this issue by using the Fisher matrix
  approximation and taking analytic derivatives around $\fnl=0$ (and a few
  other values), which enabled us to only evaluate the covariance at the
  fiducial Gaussian model.  With real data, however, a full exploration of
  parameter space will be necessary, which might be sufficiently time
  consuming to warrant analysis using a smaller set of observable
  parameters. For example, one could resort to using larger pixels and a
  coarser binning in redshift, or perhaps using no pixels at all.  One could
  also explore speeding up the covariance calculations with various
  mathematical tricks.

\item {\it Mass function:} we have assumed the \citet{Dalal} mass function
  which has been calibrated from numerical simulations and simply shifts the
  mass of halos with non-Gaussianity. A number of alternative mass functions
  have recently been proposed in the literature and studied numerically
  \cite{LoVerde,Grossi}. While the agreement in the relevant quantity $n_{\rm
    NG}(M, z)/n_{\rm G}(M, z)$ is becoming good, there is still no uniform
  agreement in the community about the convergence.  The overall constraints
  are expected to be robust given that most of the effect of non-Gaussianity
  comes from the bias scaling as $\fnl k^{-2}$ and not the mass
  function. Nevertheless, we expect constraints in this paper to be on the
  conservative side: given that the Dalal et al.\ mass function predicts a
  smaller effect due to non-Gaussianity than some of the other popular
  functions, use of these other mass functions would only increase the effects
  due to non-Gaussianity and thus improve the error bars on $\fnl$.

\item {\it Corrections to the bias formula:} While the dependence of bias on
  $\fnl$ is established to follow Eq.~(\ref{eq:bias}) both analytically and
  numerically, it could be that there are second-order corrections to the bias
  formula. These have been discussed in the literature; for example, it
  appears that a small constant offset in bias is warranted by the simulations
  and some analytical results \cite{Desjacques_Seljak_Iliev,PPH,GP}. 
  Study of  these higher-order corrections is very important but, given that 
  there is no convergence in the community on this issue as of yet, we leave their
  inclusion for future work.

\item {\it Relativistic corrections and gauge dependence:}
  \citet{wands_slosar} have shown that, to first-order, the scale-dependent
  bias does not receive relativistic corrections at large scales, using a
  spherical collapse model.  However, other authors have shown that
  higher-order corrections in the matter perturbations can produce 
non-Gaussianity (see  e.g.\ \cite{Fitzpatrick,Bartolo}).  
How the higher-order correction propagate to the halo bias is yet to be 
understood in detail.

\item {\it Observational systematics:} In this paper we have modeled the
  systematic uncertainties in understanding of the Gaussian bias $b_0(M, z)$
  and the relation between cluster mass and its observational proxy by
  introducing nuisance parameters that describe uncertainty in these
  relations. However, we have not attempted to model observational
  uncertainties, such as variations in atmospheric seeing or photometric
  calibration. Clearly, knowledge of such uncertainties over large angular
  scales will be important if measurements of non-Gaussianity are not to be
  substantially degraded. 
  We leave the study of observational systematics for
  future work.
\end{itemize}

\section{Conclusions}\label{sec:concl}

In this paper we studied how well primordial non-Gaussianity of the
local type can be probed with galaxy clusters. We took into account cluster
number counts, as well as the full covariance of cluster counts-in-cells. We
allowed generous uncertainties in the knowledge of the cluster mass-observable
relation, the photometric redshifts, and the Gaussian halo bias (we did not
consider systematics due to uncertainties in angular selection, which may be
important.)
As we discuss at length in Sec.~\ref{sec:calc}, the Fisher matrix calculation is computationally
challenging, and we resorted to a number of conservative approximations,
the most important of which is using very large pixels. 
Since angular selection issues are expected to be most significant at small angular 
scales, our pixel choices partly justify neglecting angular uncertainties.

We found that most information on primordial non-Gaussianity comes from the
previously neglected covariance of counts. The covariance links cluster
overdensities across large distances, and thus benefits the constraints on
primordial non-Gaussianity of the local type. The reason is easy to
understand: the non-Gaussian parameter $\fnl$ enters through the term
proportional to $k^{-2}$ in the bias, and correlates cluster counts in bins
separated by hundreds of megaparsecs. Other cosmological parameters do not
lead to these far-separation correlations in cluster counts (see the right
panel of Fig.~\ref{fig:Sij}). Correlations of cluster counts across vast
spatial distances of hundreds of megaparsecs therefore represent a
smoking-gun signature of primordial non-Gaussianity of the local type.

The combination of counts and clustering is particularly effective at breaking
degeneracies of $\fnl$ with cosmological and nuisance parameters, since the
two statistical probes complement each other very well. While our full set of
23 freely varying nuisance parameters can degrade $\fnl$ constraints by
factors of a few, even modest prior uncertainties on some of them break
degeneracies and restore the accuracy in $\fnl$.
For example, the bias in each photo-z bin needs to be known to 0.01 to keep
$\fnl$ constraints within $15\%$ of their values for the case of perfectly
known photo-z's.

We investigated the sensitivity of our results to the choice of fiducial value
of $\fnl$ and found that the uncertainty in $\fnl$ at $\fnl \neq 0$ is smaller
than that for $\fnl=0$. In other words, a non-zero small value of $\fnl$ may
even be more sensitively differentiated from the $\fnl=0$ case than indicated
in our Tables; the reason for this is explained in
Sec.~\ref{sec:disc_fiducial}.

Our forecasts indicate very strong constraints on primordial non-Gaussianity,
which is perhaps surprising. However, closer inspection reveals a number of
effects that help clusters achieve these numbers; we discuss these in
Sec.~\ref{sec:clus_gal}. In particular, we use the pixel-based estimator,
which is well suited for extracting signal from very large scales. Previous
error forecasts of non-Gaussianity from galaxy clustering used the suboptimal
FKP estimator; dark-energy studies that did use the pixel-based estimator
only considered variance of cluster counts.

To achieve the full potential of forecasted constraints discussed here, a few
more issues need to be carefully studied. Particularly important are
theoretical uncertainties in linking dark matter halos to observed clusters of
galaxies, and observational systematics across large angular
scales. While constraints on primordial non-Gaussianity have improved two
orders of magnitude between COBE \cite{Komatsu_thesis} and WMAP \cite{wmap7},
another one or even two orders of magnitude improvement may be possible with
upcoming surveys of large-scale structure, especially if they include both
dark matter halo counts and their clustering covariance.


\section*{Acknowledgements}
We are extremely grateful to Neal Dalal for contributing crucially to this
project at its early stages.  We thank Wayne Hu for pointing out the reference
\cite{Tegmark_etal} to us, and An\v{z}e Slosar and Adam Becker for useful
discussions.  We also thank the Aspen Center for Physics, where this work
started, for hospitality.  CC and DH are supported by the DOE OJI grant under
contract DE-FG02-95ER40899. DH is additionally supported by NSF under contract
AST-0807564, and NASA under contract NNX09AC89G.  Part of the research
described in this paper was carried out at the Jet Propulsion Laboratory,
California Institute of Technology, under a contract with the National
Aeronautics and Space Administration.

\appendix
\section{Parametrization of mass-observable relation} \label{app:mobs}

We assume a log-normal form for the probability of measuring an observable 
signal, denoted $\Mobs$, given true mass $M$, 

\begin{equation}
p(\Mobs | M) = \frac{1}{ \sqrt{2\pi} \siglnM}  \exp\left[ -x^2(\Mobs) \right] ,
\end{equation}

\noindent where

\begin{equation}
x(\Mobs) \equiv \frac{ \ln \Mobs - \ln M - \ln  \Mbias(\Mobs,z)}{ \sqrt{2} \siglnM(\Mobs,z)}.
\label{eqn:x1mobs}
\end{equation}

For the optical survey, the mass threshold of the observable is set to
$\Mth=10^{13.7}h^{-1}\Msun$ and the redshift limit is $z=1$, corresponding to
the projected sensitivity of the Dark Energy Survey.  Different studies
suggest a wide range of scatter for optical observables, ranging from a
constant $\siglnM=0.5$ \citep{wu08} to a mass-dependent scatter in the range
$0.75 < \siglnM < 1.2$ \citep{bec07}.  Using weak lensing and X-ray analysis
of MaxBCG selected optical clusters, Ref.~\cite{roz08a} estimated a lognormal
scatter of $\sim 0.45$ for $P(M|\Mobs)$, where $M$ was determined using weak
lensing and $\Mobs$ was an optical richness estimate.
We choose a fiducial mass scatter of $\siglnM=0.5$ and allow for a cubic
evolution in redshift and mass:
\begin{eqnarray}
{\rm ln}\Mbias(\Mobs,z)&=&{\rm ln}\Mbias_0 + a_1\ln(1+z)\nonumber \\[0.2cm]
&+&a_2({\rm ln}\Mobs-{\rm ln}M_{\rm pivot}), \label{eqn:mbiasdef}\\[0.0cm]
\siglnM^2(\Mobs,z)&=&\sigma_{0}^2 + \sum_{i=1}^{3}b_iz^i \nonumber \\[-0.2cm]
&+& \sum_{i=1}^{3}c_i({\rm ln}\Mobs-{\rm ln}M_{\rm pivot})^i. \label{eqn:msigdef}
\end{eqnarray}
\noindent We set $M_{\rm pivot}=10^{15}h^{-1} \Msun$. 
In all, we have 10 nuisance parameters for the optical mass errors 
(${\rm ln}\Mbias_0$, $a_1$, $a_2$, $\sigma_{0}^2$, $b_i$, $c_i$).

There are few, if any, constraints on the number of parameters necessary to
realistically describe the evolution of the variance and bias with mass.
Ref.\ \cite{lim05} shows that a cubic evolution of the mass-scatter with
redshift captures most of the residual uncertainty when the redshift evolution
is completely free (as assumed in the Dark Energy Task Force (DETF) report
\cite{DETF}).
 While generous, this parametrization assumes a lognormal distribution of 
the mass-observable relation that may fail for low-masses (see e.g. \cite{coh09}).
However, \cite{clusters_PC} show that more complex distributions do not degrade
results substantially ($\sim20-30\%$ for the test case assumed by the authors). 
We have also implicitly assumed that selection effects can be described by the
bias and scatter of the mass-observable relation.  By the year 2016, we expect
significant progress in simulations of cluster surveys that will allow us to
better parametrize the cluster selection errors.

\section{Photometric redshift errors and Gaussian halo bias} \label{app:photoz}

Uncertainties in the redshifts distort the volume element.
Assuming photometric techniques are used to determine the redshifts of 
the clusters (hereafter photo-z's), and a perfect angular selection the mean number of clusters 
in a photo-z bin $\zphot_i \le \zphot \le \zphot_{i+1}$ is

\begin{eqnarray}
\bar m_{\alpha,i} &=& \int_{\zphot_i}^{\zphot_{i+1}} d\zphot 
\int dV \bar n_{\alpha} W_{i}^{\rm th}(\Omega) p(\zphot| z)
\label{eqn:numwin}
\end{eqnarray}

\noindent where $W^{\rm th}_{i}(\Omega)$ is an angular top hat window function.
We parametrize the probability of measuring a photometric
redshift, $\zphot$, given the true cluster redshift $z$ as \cite{lim07}  
\begin{eqnarray}
p(\zphot| z) &=& \frac{1}{\sqrt{2\pi \sigz^2}} \exp\left[ -y^2(\zphot) \right]
\label{eqn:pzpzs}
\end{eqnarray}
\noindent where
\begin{eqnarray}
y(\zphot)&\equiv& \frac{\zphot -z -\zbias}{ \sqrt{2\sigz^2}}, 
\end{eqnarray}

\noindent $\zbias$ is the photometric redshift bias and $\sigz$ is the scatter in the 
photo-z's.

On large scales, the number counts of clusters $m({\bf{x}})$ trace
the linear density perturbation $\delta({\bf{x}})$
\begin{equation}
m_i(M_\alpha, {\bf x}) \equiv m_{i\alpha}= \bar{m}_i ( 1 + b(M_\alpha, z)\delta({\bf x}))
\end{equation}
where $M_\alpha$ denotes a bin in mass and $i$ refers to the pixel on the
sky defined by its angular location and redshift. 
The (Gaussian) halo bias may be very roughly approximated by
\citep{ShethTormen99}
\begin{equation}
b_0(M;z) = 1 + \frac{a_c \delta_c^2/\sigma^2 -1}{\delta_c} 
         + \frac{ 2 p_c }{ \delta_c [ 1 + (a \delta_c^2/\sigma^2)^{p_c}]}
\label{eqn:bias}
\end{equation}
\noindent with $a_c=0.75$, $p_c= 0.3$, and $\delta_c=1.69$.  Here
$\sigma(M,z)$ is the amplitude of mass fluctuations on scale $M$, defined as
  usual by
\begin{equation}
\sigma^2 = \int \frac{k^3}{2\pi^2} P(k)\,W^2(kR) \frac{dk}{k},
\label{eq:sigma}
\end{equation}
where $W(x)=3 j_1(x)/x$ (the top-hat window), $R=(3M/4\pi\bar\rho_m)^{1/3}$,
and $P(k)$ and $\bar\rho_m$ are the matter power spectrum and energy density
respectively.

Integrating the expression above yields the average cluster linear bias:
\begin{eqnarray}
b_{\alpha,i}(z) &=& \frac{1}{\bar n_{\alpha,i}(z)}  
\int_{\Mobs^{\alpha}}^{\Mobs^{\alpha+1}} \frac{d{\Mobs}}{\Mobs}\int \frac{d M}{M} \nonumber \\
&&\times \frac{d \bar n_{\alpha,i}(z)}{d\ln M} b(M;z)p(\Mobs|M).
\label{eqn:linbias}
\end{eqnarray}

\bibliography{fnl}

\end{document}